\documentclass[preprint,12pt]{elsarticle}

\usepackage{amssymb}

\journal{Physics Letter B}

\begin{document}

\begin{frontmatter}



\title{Covariance Analysis of Symmetry Energy Observables from Heavy Ion Collision}


\author[1]{Yingxun Zhang\corref{cor1}}
\ead{zhyx@ciae.ac.cn}
\cortext[cor1]{Corresponding author}
\author[2] {M.B.Tsang} 
\author[1] {Zhuxia Li} 

\address[1] {China Institute of Atomic Energy, P.O. Box 275 (10), Beijing 102413, P.R. China}
\address[2] {National Superconducting Cyclotron Laboratory, Michigan State University, East Lansing, MI 48824, USA}

\begin{abstract}
Using covariance analysis, we quantify the correlations between the interaction parameters in a transport model and the observables commonly used to extract information of the Equation of State of Asymmetric Nuclear Matter in experiments. By simulating $^{124}$Sn+$^{124}$Sn, $^{124}$Sn+$^{112}$Sn and $^{112}$Sn+$^{112}$Sn reactions at beam energies of 50 and 120 MeV per nucleon, we have identified that the nucleon effective mass splitting are most strongly correlated to the neutrons and protons yield ratios with high kinetic energy from central collisions especially at high incident energy. The best observable to determine the slope of the symmetry energy, L, at saturation density is the isospin diffusion observable even though the correlation is not very strong ($\sim$0.7). Similar magnitude of correlation but opposite in sign exists for isospin diffusion and nucleon isoscalar effective mass. At 120 MeV/u, the effective mass splitting and the isoscalar effective mass also have opposite correlation for the double n/p and isoscaling p/p yield ratios. By combining data and simulations at different beam energies, it should be possible to place constraints on the slope of symmetry energy (L) and effective mass splitting with reasonable uncertainties.
\end{abstract}


\end{frontmatter}


Knowledge about the isospin asymmetric nuclear equation of state (EoS) is of fundamental importance to our understanding of nature's most asymmetric objects including neutron stars and heavy nuclei composed of very different numbers of neutrons and protons. Theoretically, there are two microscopic approaches to describe the EoS of nuclear matter.
One approach starts from a realistic two-body free NN interactions\cite{Wiringa95,Mach01} as input to the relativistic Dirac-Bruckner-Hartree-Fock (DBHF) and its nonrelativistic counterpart Bruckner-Hartree-Fock (BHF)\cite{teHaar87,Cugnon87,Baldo00,WZuo02,Dalen05} and chiral effective field theory\cite{Hebeler10}. Another one is to use effective density-dependent many-body interactions such as the zero-range Skyrme interaction\cite{Skyrme56,Vauthe72,Chab97}, finite-range Gogny interaction\cite{Decharge80} and effective Lagrangian\cite{Brock78,Horo83,Bouy87,WHLong06} as inputs leading to Skyrme-Hartree-Fock (SHF) \cite{Vauthe72,Chab97}, Gogny-Hartree-Fock\cite{Blaiz95} and Relativistic-Hartree-Fock (RHF) approaches\cite{Brock78,Horo83, Bouy87,WHLong06}.
Of all interactions, the effective Skyrme interaction are more commonly used in nuclear structure, reactions and astrophysics studies as the effective Skyrme interactions are relatively simple mathematically to make it computationally feasible\cite{Greiner96} and contain sufficient physics to allow quantitative description of heavy nuclei. Furthermore, the Fock term in the non-relativistic SHF approach can be managed computationally while the Fock terms in Relativistic HF approaches are difficult to compute.
The interactions parameters are usually obtained by fitting the properties of symmetric nuclear matter (such as the saturation density and the corresponding energy per nucleon and its incompressibility), properties of asymmetric nuclear matter (such as symmetry energy and isovector effective mass at saturation density), finite nuclei properties (such as binding energies and r.m.s radius of selected set of doubly magic nuclei) etc.\cite{Chab97}. In Ref \cite{Dutra12}, 240 Skyrme parameter sets that fit the ground-state properties of stable nuclei, symmetric and asymmetric nuclear matter are compiled. The large number of parameterization arises in part because there are strong correlations between individual parameters or groups of parameters, that fit particular physical properties of the many-body nuclear system. These sets lead to very different Equation of States of pure neutron matter\cite{Brown00,BALi08,Tsang12} which may have different incompressibility $K_0=9\rho^2 \frac{\partial^2 \epsilon/\rho}{\partial \rho^2}$, symmetry energy coefficient $S_0=S(\rho_0)$, slope of symmetry energy $L=3\rho_0\frac{\partial S(\rho)}{\partial \rho}|_{\rho_0}$,  isoscalar effective mass $\frac{m}{m_s^*}=(1+\frac{2m}{\hbar^2}\frac{\partial}{\partial \tau}\frac{E}{A})|_{\rho_0}$\cite{Klu09}, and isovector effective mass $m_v^*=\frac{1}{1+\kappa}$, where $\kappa$ is the enhancement factor of the Thomas-Reich-Kuhn sum rule \cite{PRing80}. In this work, the magnitude of effective mass splitting, $(m_n^*-m_p^*)/m$  was not used as input variables since its form is much more complicated to be incorporated into the code. Instead we use $f_I=\frac{1}{2\delta}(\frac{m}{m_n^*}-\frac{m}{m_p^*})=\frac{m}{m_s^*}-\frac{m}{m_v^*}$, where $\delta=(\rho_n-\rho_p)/(\rho_n+\rho_p)$, $\rho_n$ and $\rho_p$ are the neutron and proton density, $m_n^*$, $m_p^*$ and $m$ are the neutron, proton effective mass and free nucleon mass. In Skyrme Hartree-Fock approximation\cite{Vauthe72,Chab97,Dutra12}, $f_I$ increases with increasing density, but is independent of the momentum and kinetic energy. In the DBHF and RHF approximations \cite{WHLong06,WZuo05,RChen12}, $f_I$ not only depends on the density but also on the kinetic energy of the in-medium nucleon.

Some properties of symmetric nuclear matter, such as $K_0=230\pm30$MeV\cite{Blaiz95,Dutra12,Young99} and $m_s^*/m=0.65-0.9$\cite{Blaiz95,Dutra12,Klu09,Danie00,Chap08,Bohi79}, have been extracted from isoscalar collective vibrations, giant quadrupole resonance and heavy ion collisions measurements. In addition, constraints on $E_0(\rho)$ and pressure $P$ have been obtained in density regions ranging from saturation density to five times normal densities using collective flow and kaon production data in energetic nucleus-nucleus collisions\cite{Danie02, Fuch06,Lynch09}.
To obtain information of the symmetry energy with heavy ion collision data, the symmetry potential used in transport models is changed by varying its input parameters, corresponding to the different values of $S_0$  and/or $L$ in the expression of density dependence of symmetry energy. The results of the calculations are then compared with data to find the best parameter sets. Recently, a consistent set of constraints on the symmetry energy near saturation density between $S_0$, and its slope, $L$, has been obtained from observables measured in both nuclear structure and nuclear reaction experiments\cite{Tsang12,Tsang09,Latti14,Horowitz14}.

In Skyrme parametrizations, the symmetry energy are correlated to other parameters such as those associated with the nucleon effective mass $m_s^*$ and isovector effective mass $m_v^*$.
Such correlations would affect the uncertainties of symmetry energy constraints obtained from heavy ion collision data. In order to achieve the goal of obtaining precise and accurate symmetry energy constraints, one has to identify what experimental observables are crucial for better constraining the interested physical quantities in the theoretical models\cite{Ire15}.
Ideally, one would do a global chi-square analysis using existing data to obtain the best set of model parameters. Then a covariance matrix can be obtained between any two model parameters using a chi-square fit\cite{Sieg}.
Currently, this is not feasible considering the intensive CPU time needed to do transport model calculations. As a start to tackle this issue, we propose to use 12 parameters sets to perform covariance analysis to quantitatively examine the correlations between model parameters A and observables B commonly used in experiments.
The linear-correlation coefficient $C_{AB}$  between variable $A$ and observable $B$ is calculated as follows\cite{Sieg}:
\begin{eqnarray}
C_{AB}&=&\frac{cov(A,B)}{\sigma(A)\sigma(B)}\\
cov(A,B)&=&\frac{1}{N-1}\sum_{i}(A_i-<A>)(B_i-<B>)\\
\sigma(X)&=&\sqrt{\frac{1}{N-1}\sum_i(X_i-<X>)^2},      X=A, B\\
<X>&=&\frac{1}{N}\sum_i X_i , i=1, N.
\end{eqnarray}
$cov(A,B)$ is the covariance, $\sigma(X)$ is the variance. $C_{AB}=\pm1$ means there is a linear dependence between $A$ and $B$, and $C_{AB}$ = 0 means no correlations.

We use the Improved Quantum Molecular Dynamic Model which incorporates the effective Skyrme interactions (ImQMD-Sky)\cite{Zhang14} to simulate the collisions of heavy ions with parameter sets listed in Table I. $A_i$ represents the ith parameter set of the transport variable A where A=$K_0$, $S_0$, $L$, $m_s^*$, $m_v^*$ or $f_I$ used as input to the ImQMD-Sky. 
In ImQMD-Sky, the nucleonic potential energy density is $u_{loc}+u_{md}$, where
\begin{eqnarray}
u_{\rho}&=&\frac{\alpha}{2}\frac{\rho^2}{\rho_0}+
\frac{\beta}{\eta+1}\frac{\rho^{\eta+1}}{\rho_0^{\eta}}+
\frac{g_{sur}}{2\rho_0}(\nabla\rho)^2\nonumber\\
&  &+\frac{g_{sur,iso}}{\rho_0}[\nabla(\rho_n-\rho_p)]^2\nonumber\\
&  &+A_{sym}\rho^2\delta^2
+B_{sym}\rho^{\eta+1}\delta^2
\end{eqnarray}
and the energy density of Skyrme-type momentum dependent interaction are written based on its interaction form $\delta(r_1-r_2)(p_1-p_2 )^2$ \cite{Skyrme56,Vauthe72},
\begin{eqnarray}
u_{md} &=& u_{md}(\rho\tau)+u_{md}(\rho_n\tau_n)+u_{md}(\rho_p\tau_p)\\\nonumber
&=& C_0 \int d^3p d^3p' f(\vec r,\vec p)f(\vec r,\vec p')(\vec p-\vec p')^2 +\\\nonumber
& &D_0 \int d^3p d^3p' [f_n(\vec r,\vec p)f_n(\vec r,\vec p')(\vec p-\vec p')^2\\\nonumber
&  & +f_p(\vec r,\vec p)f_p(\vec r,\vec p')(\vec p-\vec p')^2].
\end{eqnarray}
The 9 parameters $\alpha$, $\beta$, $\eta$, $A_{sym}$, $B_{sym}$, $C_0$, $D_0$, $g_{sur}$, $g_{sur,iso}$ used in ImQMD-Sky can be derived from standard Skyrme parameter sets with 9 parameters \{$t_0$, $t_1$, $t_2$, $t_3$, $x_0$, $x_1$, $x_2$, $x_3$, $\sigma$\}\cite{Zhang14,Zhang06}. The coefficients of the surface terms are set as $g_{sur}=24.5MeVfm^2$ and $g_{sur;iso}= -4.99MeVfm^2$ which are the same values derived from SLy4 parameter set\cite{Chab97}. Varying $g_{sur}$ and $g_{sur,iso}$ for different Skyrme interactions have negligible effects on the experimental observables at intermediate energy. By using the relationship which was derived in reference\cite{Agra05,LWChen09}, the reduced 7 Skyrme parameter sets \{$\alpha$, $\beta$, $\eta$, $A_{sym}$, $B_{sym}$, $C_0$, $D_0$\} can be replaced by the parameter sets \{$\rho_0$, $E_0$, $K_0$, $S_0$, $L$, $m_s^*$, $m_v^*$\}  which is directly related to the properties of nuclear matter at saturation density $\rho_0$. Choosing the experimental values of $\rho_0=0.16fm^{-3}$, $E_0=-16MeV$, the parameter sets are further reduced to 5 variables, A=$K_0$,$S_0$,$L$,$m_s^*$,$m_v^*$. As explained above, to simplify the coding, the input variables used in ImQMD-Sky are A=$K_0$,$S_0$,$L$,$m_s^*$,$f_I(m_s^*,m_v^*)$.

For HIC observables, we adopt the ratios constructed from nucleon spectra.
Most transport models cannot describe accurately the absolute yield of free nucleons due to models inadequacies in describing light clusters\cite{Neuba99,Ono12,Zhang12R}. This problem can be largely alleviated by constructing "coalescence invariant" quantities, i.e., nucleon observables summed over all light clusters, which show much better agreement between theory and experiment\cite{Horowitz14,Famiano06,Zhang08,Coupland14}. We construct the coalescence invariant (CI) nucleon yield spectra and their ratios in the same way as in Ref\cite{Famiano06,Zhang08,Coupland14} by combining the nucleons in the light particles and free nucleons at given kinetic energy per nucleon as follows,
\begin{eqnarray}
\frac{dY_i(n)}{dE_{c.m.}/A}=\sum N \frac{dM_i(N,Z)}{dE_{c.m.}/A}\\
\frac{dY_i(p)}{dE_{c.m.}/A}=\sum Z \frac{dM_i(N,Z)}{dE_{c.m.}/A}
\end{eqnarray}
The summation is up to $Z\le6$ and $A\le16$ particles in the calculations, $\frac{dM_i(N,Z)}{dE_{c.m.}/A}$ is the multiplicity of fragments with neutron number N and proton number Z at certain kinetic energy. Such CI neutron and proton yields obtained from ImQMD-Sky calculations with selected Skyrme parameter sets reproduce the Sn+Sn data\cite{Coupland14} reasonably well especially at high beam energy. For simplicity, integrated CI neutron and proton yields from reaction $i$ are represented as $Y_i(n)$ and $Y_i(p)$ respectively in the following. By convention, the more neutron-rich reaction is represented by the subscript "2" in this work, i=1,2 represent the reactions $^{112}$Sn+$^{112}$Sn and $^{124}$Sn+$^{124}$Sn.

We simulate 10,000 events for each of the three systems, $^{124}$Sn+$^{124}$Sn, $^{124}$Sn+$^{112}$Sn and $^{112}$Sn+$^{112}$Sn.
Four yield ratios are constructed by integrating the CI nucleon energy spectra with $E_{c.m.}/A>40MeV$. High energy nucleons are less influenced by sequential decay and are more sensitive to the effective mass splitting\cite{Zhang14,Rizzo05,Giord10,Hermann,ZQFeng12,Xie13}. The angular cut,  $70^\circ<\theta_{c.m.}<110^\circ$, is also imposed to reduce the contributions to the CI nucleon spectra from reaction residues.
The four ratio observables are: 1.) single n/p ratio CI-$R_2(n/p)$=$Y_2(n)$/$Y_2(p)$, 2.) double n/p ratios CI-DR(n/p)=CI-R$_2$(n/p)/CI-R$_1$(n/p)=CI-R$_{21}$(n/n)/CI-R$_{21}$(p/p), 3.) isoscaling ratios CI-R$_{21}$(n/n)=Y$_2$(n)/Y$_1$(n) and 4.) CI-R$_{21}$(p/p)=Y$_2$(p)/Y$_1$(p) ratios. The experimental isoscaling ratios and the double ratios have the advantage that they are minimally affected by detector systematic uncertainties and sequential decays\cite{Tsang01,Chajecki}. 5.) the fifth and last isospin observable we examine is the isospin transport ratios\cite{Tsang04}, $R_{diff}$, that quantify diffusion of the nucleons in the neck region during the nuclear collisions. $R_{diff}$ have been used to constrain the symmetry energy at subsaturation density. These constraints are consistent with constraints obtained in nuclear structure experiments\cite{Tsang12}.

The isospin diffusion observable, $R_{diff}$ is defined as \cite{Tsang04}, $R_{diff}=(2X-X_{aa}-X_{bb})/(X_{aa}-X_{bb})$
where $X=X_{ab}$ is an isospin observable. In this work, we use the subscripts $a$ and $b$ to denote the projectile (first index) and target (second index) combination. By convention, $aa$ and $bb$ represent the neutron-rich and neutron-poor reactions, respectively. In theoretical calculations, $X$ is the isospin asymmetry of the emitting source, which is constructed from all the emitted nucleons and fragments with velocity cut $v>0.5v_{beam}^{c.m.}$. It is assumed to be linearly related to the experimental observable $X_{exp}$ as discussed in \cite{Tsang04,Zhang12}. $R_{diff}$ are constructed with at least three reaction systems to cancel the drift and retain the information of the diffusion. Since the changes of isospin asymmetry of emitted source mainly come from the nucleon diffusion between projectile and target, $R_{diff}$ carry information of "nucleons" that have energy lower than 40MeV. Thus complimentary information is obtained from the nucleon ratios with high energy gate and the isospin transport ratios.

We first perform calculations by separately varying each variable of the parameter set $K_0$, $S_0$, $L$, $m_s^*$, $f_I$. In the following studies, 12 parameter sets are constructed and listed in Table I. The average values of each variable in the parameter space we used are, $<K_0>=242.5 MeV$, $<S_0>=32MeV$, $<L>=54.5MeV$, $<m_s^*/m>=0.7375$ and $<f_I>=-0.1835$.
Unlike refers.\cite{Tsang09,Cozma13}, there is no attempt to optimize interaction parameters to fit the experimental observables in this work. Deviation of $<A>$ and $<B>$ values from their "experimental" values is one source of uncertainties in the calculated correlation coefficients $C_{AB}$. As it is difficult to track the systematic uncertainties in the transport model calculations, we do not estimate uncertainties in this work.

\begin{table}[htbp]
\caption{\label{tab:table1}%
List of twelve parameter sets used in the ImQMD calculations. $\rho_0=0.16fm^{-3}$, $E_0=-16MeV$, and $g_{sur}=24.5MeVfm^2$, $g_{sur,iso}=-4.99MeVfm^2$}
\begin{tabular}{lccccc}
\hline
\hline
\textrm{Para.}&
\textrm{$K_0$ (MeV)} & \textrm{$S_0$(MeV)} & \textrm{$L$ (MeV)} & \textrm{$m^*_s/m$} &
$f_I$\\
\hline
 1 &  \textbf{230} & \textbf{32} & \textbf{46} & \textbf{0.7} & \textbf{-0.238} \\
 2 &  \textbf{280} & 32 & 46 & 0.7 & -0.238 \\
 3 &  \textbf{330} & 32 & 46 & 0.7 & -0.238 \\
 4 &  230 & \textbf{30} & 46 & 0.7 & -0.238 \\
 5 &  230 & \textbf{34} & 46 & 0.7 & -0.238 \\
 6 &  230 & 32 & \textbf{60} & 0.7 & -0.238 \\
 7 &  230 & 32 & \textbf{80} & 0.7 & -0.238 \\
 8 &  230 & 32 & \textbf{100} & 0.7 & -0.238 \\
 9 &  230 & 32 & 46 & \textbf{0.85} & -0.238 \\
 10 &  230 & 32 & 46 & \textbf{1.00} & -0.238 \\
 11 &  230 & 32 & 46 & 0.7 & \textbf{0.0} \\
 12(SLy4) &  230 & 32 & 46 & 0.7 & \textbf{0.178} \\
 \hline
\end{tabular}
\end{table}

To study the sensitivity of different force parameters in Skyme interactions on isospin observables, we calculate the covariance coefficients $C_{AB}$ based on Eq.(1)-(4) between five force parameters: A=$K_0$, $S_0$, $L$, $m_s^*$ and $f_I$ and five HIC observables: B=CI-$R_2(n/p)$, CI-DR(n/p), CI-$R_{21}(n/n)$, CI-$R_{21}(p/p)$ and $R_{diff}$. All the nucleon yield observables in the simulations are obtained for $^{124}$Sn+$^{124}$Sn and $^{112}$Sn+$^{112}$Sn collisions at incident energies of 50 and 120 MeV/u at impact parameter 2fm. For $R_{diff}$, an additional mixed reaction $^{124}$Sn+$^{112}$Sn is included and the calculations are performed with mid-peripheral impact parameter of 6 fm where a low density neck is formed and isospin transport between projectile and target occurs.

Figure 1 shows the correlation coefficient $C_{AB}$ for Sn+Sn reactions at $E_{beam}$=50 MeV/u (upper panels) and at  $E_{beam}$=120 MeV/u (lower panels). Red color bars represent positive correlations which mean observable B increases with parameter A, and blue color bars show negative correlations which mean observable B decreases with increasing parameter A. To focus our search for the best experimental observables which are most sensitive to the model parameters, we will discuss only $C_{AB}$ with values greater than 0.6, which were represented as solid blue and red bars in Figure 1. With this criterion, $S_0$ and $K_0$ are not very sensitive to any of the observables studied here.

\begin{figure}[htbp]
\centering
\includegraphics[angle=270,scale=0.4]{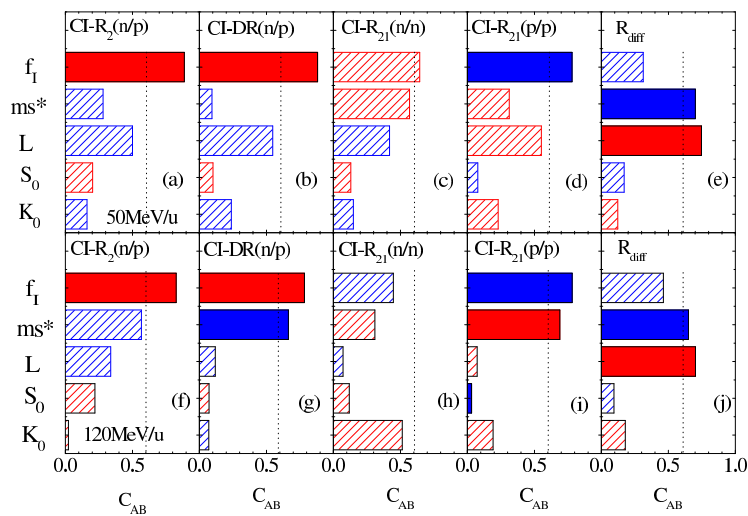}
\setlength{\abovecaptionskip}{50pt}
\caption{(Color online)
Correlations of five obervables, CI-$R_2(n/p)$ (a), CI-$DR(n/p)$ (b), CI-$R_{21}(n/n)$ (c), CI-$R_{21}(p/p)$ (d), $R_{diff}$ (e) with five force parameter, $K_0$, $S_0$, $L$, $m_s^*$ and $f_I$. Up panels are the results for 50 MeV per nucleon, and bottom panels are for 120 MeV per nucleon.}
\setlength{\belowcaptionskip}{0pt}
\end{figure}
In the case of L, the slope of the symmetry energy at saturation density, $C_{L,Rdiff}\sim0.7$ meets this criterion. Our analysis also shows negative correlations with $m_s^*$, $C_{m_s^*,R_{diff}}\sim-0.64$ to $-0.70$. The observation that $R_{diff}$ is also sensitive to $m_s^*$ is consistent with the results from BUU calculations\cite{LWChen05,Rizzo08,Coupland11}, where $R_{diff}$ values decrease with increasing $m_s^*$.
The relatively strong correlations exhibited by this observable with $m_s^*$ and $L$ should allow one to extract the constraints of both parameters simultaneously with reasonable uncertainties, or extract a range of $L$ values using $m_s^*=0.65-0.9$ obtained in ref.\cite{Blaiz95,Danie00,Dutra12,Klu09,Chap08,Bohi79}.
In ref.\cite{Zhang14,Rizzo05,Giord10,Hermann,ZQFeng12,Xie13}, the low energy free particle energy spectra show sensitivity to the slope of symmetry energy while high energy particle spectra is sensitive to the effective mass splitting. When the energy cut of $E_{c.m.}/A<40 MeV$ is applied, the correlation between L and CI-$R_2(n/p)$; CI-DR(n/p) and CI-$R_{21}(p/p)$ is strong. This quantity will be examined in more details when the correlation matrix is compared to data in a future study.


$f_I$ shows the strongest correlations to the experimental coalescence invariant nucleon observables, the single and double n/p yield ratios at both incident energies. As shown in panels (a), (f) and (b), (g) of Figure 1, the correlation between the single and double n/p yield ratios and $f_I$ is larger than 0.8. The positive correlation means that the results obtained with $m_n^*<m_p^*$ are greater than that with $m_n^*>m_p^*$ which is consistent with the previous published results from other transport model calculations\cite{Rizzo05,Giord10,Hermann,ZQFeng12,Xie13}. Strong correlation with $m_s^*$ is observed mainly at high energy, because the more violent nucleon-nucleon collisions at 120MeV/u cause larger momentum transfer and lead to the momentum dependent interaction (which can be characterized by the effective mass) to play more important roles. This can be understood from the expression of
the density dependence of symmetry energy in Skyrme interaction, where $S(\rho)$ not only depends on isovector effective mass $m_v^*$, but also on the isoscalar effective mass $m_s^*$, i.e. $S(\rho)=\frac{1}{3}\epsilon_F\rho^{2/3}+A_{sym}\rho+B_{sym}\rho^{\sigma+1}+C_{sym} (m_s^*,m_v^* ) \rho^{5/3}$.

For isoscaling ratios, the sensitivity of CI-$R_{21}(n/n)$ to all variables $K_0$, $S_0$, $L$, $m_s^*$ and $f_I$ are borderline. On the other hand, CI-$R_{21}(p/p)$ are particularly sensitive to $f_I$ due to the Coulomb repulsion as protons accelerate to higher kinetic energy than neutrons. Since CI-$R_{21}(p/p)$ depends on the difference of chemical potential for proton between system "2" and "1", i.e. $\mu_{p2}-\mu_{p1}$ which is equal to $-V_{sym}(\delta_2-\delta_1)=-\int dE_k f_I \Delta\delta=-\int dE_k(\frac{m}{m_s^*}-\frac{m}{m_v^*})\Delta\delta$, where $V_{sym}$ is the symmetry potential and $E_k$ is the kinetic energy of nucleon,
it leads to positive correlation to $m_s^*$ and negative correlation to $f_I$.
At higher beam energy 120MeV/u, the influence of effective mass splitting become more important. It causes the strong negative correlation between CI-$R_{21}(p/p)$ and $f_I$, and positive correlation between CI-$R_{21}(p/p)$ and $m_s^*$. One should be able to extract constraints for both $f_I$ and $m_s^*$ from the single and double n/p ratios as well as the p/p ratios at 120MeV/u incident energy. Alternately, $f_I$ can be extracted using the value of $m_s^*/m=0.65-0.9$ obtained in ref.\cite{Blaiz95,Danie00,Dutra12,Klu09,Chap08,Bohi79}.

In summary, by separately varying the interaction parameter sets in the ImQMD-Sky code, we study the influence of $K_0$,$S_0$,L,$m_s^*$,and $f_I$ on the coalescence invariant neutron and proton yield ratios at high energy region, for $^{124}$Sn+$^{124}$Sn and $^{112}$Sn+$^{112}$Sn at 50MeV/u and 120MeV/u incident energy. Sensitivities to $S_0$ and $K_0$ are relatively small, $C_{AB}<0.5$. At incident energy of 120MeV/u, strong correlations are observed between observables constructed from coalescence invariant nucleon spectra with $E_{c.m.}/A>40$ MeV, such as CI-$R_2(n/p)$, CI-DR(n/p), and CI-$R_{21}(p/p)$, and the effective mass splitting as well as the isoscalar effective mass, important input parameters to the transport models. The calculations also confirm the sensitivity of $L$ to the isospin diffusion observable. Since the same observable is also sensitive to isoscalar effective mass, this should allow one to extract the constraints of $m_s^*$ and $L$ with reasonable uncertainties. Similarly the opposite correlations of the nucleon yield ratios, such as CI-DR(n/p) and CI-R$_2$(p/p) ratios to $f_I$ and $m_s^*$, at 120 MeV/u reactions should allow one to disentangle the effects of the effective nucleon mass splitting and the isoscalar effective mass.


\textbf{Acknowledgements}
This work has been supported by the National Natural Science Foundation of China under Grants No.(11475262, 11375062,11275072), National Key Basic Research Development Program of China under Grant No. 2013CB834404. Y.Zhang thanks Dr. Yue Zhang useful discussion on the covariance analysis method. MBT acknowledges support from the USA National Science Foundation Grants No. PHY-1102511 and travel support from CUSTIPEN (China-US Theory Institute for Physics with Exotic Nuclei) under the US Department of Energy Grant No. DE-FG02-13ER42025.








\end{document}